\def\aj{{AJ}}

\def\apj{{ApJ}}
\def\apjs{{ApJS}}

\def\HalphaNII{H$\alpha$+[NII]}

\def\SII{[SII]}

\documentclass[cup5b]{caps}
\usepackage{graphicx}
\usepackage{amssymb}
\usepackage{ociwsymp1e}
\HeadText{G. A. Verdoes Kleijn, R. P. van der Marel, and J.  Noel-Storr}

\begin{document}

\pagenumbering{arabic}

\author[]{G. A. VERDOES KLEIJN$^1$, R. P. VAN DER MAREL, and J.
NOEL-STORR$^{2,3}$ \\
(1) ESO, \,\, (2) STScI, \,\, (3) Columbia University}

\chapter{What Drives the Central Velocity Dispersion in Nearby Early-Type Galaxies?}

\begin{abstract} The majority of nearby early-type galaxies contains
detectable amounts of emission-line gas at their centers. The
emission-line ratios and gas kinematics potentially form a valuable
diagnostic of the nuclear activity and gravitational potential
well. The observed central gas velocity dispersion often exceeds the
stellar velocity dispersion. This could be due to either the
gravitational potential of a black hole or turbulent shocks in the
gas. Here we try to discriminate between these two scenarios.

\end{abstract}

\section{Introduction}

Optical emission-line gas is detected in at least 50\% of the centers
of nearby early-type (E/S0) galaxies (e.g., Phillips et al. 1986;
Goudfrooij et al. 1994; Macchetto et al. 1996; Ho et al. 1997). The
gas velocity dispersion $\sigma_{\rm gas}$ (corrected for instrumental
broadening) typically increases towards the nucleus, reaching values
similar to the the central $\sigma_{\rm star}$ as measured through
100s of pc scale sized apertures (e.g., Zeilinger et al.,
1996). Closer to the nucleus, $\sigma_{\rm gas}$ often exceeds
$\sigma_{\rm star}$ (e.g., Ferrarese et al. 1996; Macchetto et al.,
1997; van der Marel \& van den Bosch 1998; Cappellari et al. 2002;
Verdoes Kleijn et al. 2002). Potential contributors to $\sigma_{\rm
gas}$ are motions due to (i) a black hole (BH) gravitational potential
or (ii) shocks/turbulence in the gas. In the former case, $\sigma_{\rm
gas}$ could be caused by three dimensional motion of collisionless
cloudlets of gas. More likely, the collisional gas is expected to have
settled in a disk. In that case, $\sigma_{\rm gas}$ will be caused by
differential rotation over the aperture. However, shocks and/or
turbulence can contribute to $\sigma_{\rm gas}$ if there is a
permanent input of kinetic energy (perhaps the active galactic
nucleus) to sustain them. Thus, determining the driver of $\sigma_{\rm
gas}$ has implications for measurements of the BH mass ($M_{\rm BH}$)
and the flow properties of potential BH accretion material. Here we
present preliminary modeling of $\sigma_{\rm gas}$ and $\sigma_{\rm
star}$ in the nuclei of 16 active and 4 quiescent early-type galaxies
to constrain the relative importance of gravitation and
shocks/turbulence.

\section{Sample and Data}

The galaxy sample consists of 16 active galaxies with radio-jets taken
from the UGC FR~I sample (see Noel-Storr et al., these proceedings, for
selection criteria) and four relatively quiescent early-type galaxies
(NGC 3078, NGC 4526, NGC 6861 and IC 1459) (Table~1). The four
quiescent early-type galaxies were selected to have central dust
and/or gas disks. Four of the 21 UGC FR~I galaxies were not included
here because either $\sigma_{\rm star}$ is not available or the
location of the nucleus could not be determined reliably, mainly due
to dust obscuration

\begin{table}
\caption{Early-Type Galaxy Sample.}
\begin{tabular}{lllll|lllll}
\hline \hline
{Galaxy} & {Type} & {$M_{\rm B}$} & {$D$} & {Galaxy} & {Type}  & {$M_{\rm B}$} & {$D$} \\
{}       & {}     & {}            & {(Mpc)} & {}       & {}      & {}            & {(Mpc)} \\
{(1)}       & {(2)}     & {(3)}            & {(4)} & {(1)}       & {(2)}      & {(3)}            & {(4)} \\
\hline
NGC 315   & E+:        & -21.9 & 67.9 & M87       & E+0-1 pec  & -21.4 & 15.4 \\
NGC 383   & SA0-:      & -21.4 & 65.2 & NGC 5127  & E pec      & -20.6 & 64.4 \\
NGC 541   & cD         & -21.2 & 73.3 & NGC 5490  & E          & -21.1 & 69.2 \\
NGC 741   & E0:        & -21.9 & 70.2 & NGC 7052  & E          & -19.9 & 55.4 \\
3C 66B    & E          & -21.1 & 84.8 & 3C 449    & S0-:       & -19.6 & 68.3 \\
NGC 2329  & S0-:       & -21.1 & 76.7 & NGC 7626  & E pec:     & -21.1 & 46.6 \\
3C 264    & E          & -21.0 & 84.4 & NGC 3078  & E2-3       & -20.6 & 35.2 \\
UGC 7115  & E          & -20.4 & 90.5 & NGC 4526  & SAB(s)0    & -20.5 & 16.9 \\
NGC 4335  & E          & -20.3 & 61.5 & NGC 6861  & SA(s)0-:   & -20.2 & 28.1 \\
M84       & E1         & -20.8 & 15.4 & IC 1459   & E3         & -21.1 & 29.2 \\
\hline \hline
\end{tabular}
The galaxies are taken from the UGC FR~I radio galaxy sample
(Noel-Storr et al., these proceedings) except for NGC 3078, NGC4526,
NGC 6861 and IC 1459. Col.(2): Hubble classification from
NED. Col.(3): Absolute blue magnitude from LEDA. Col.(4): Distances
from Faber et al.~(1989), Tonry et al.~(2001), or, if not available,
directly from recession velocity and $H_0 =
75$kms$^{-1}$Mpc$^{-1}$. Col.(5): Galaxy is part of the UGC FR~I
sample (1) or not (0).
\label{table1}
\end{table}

HST/STIS (HST/FOS for IC 1459) emission-line spectra are available for
the sample. They include the {\HalphaNII} and
\SII6716,6731 lines. A central $\sigma_{\rm gas}$ per galaxy is
determined by fitting single Gaussians to each of the lines in
{\HalphaNII} and \SII6716,6731 at the nucleus. An analytic
emission-line flux profile is determined by fitting a double
exponential to the emission-line surface brightness measurements from
the HST/STIS spectra and HST/WFPC2 emission-line images (where
available).  The central stellar dispersion $\sigma_{\rm star}$ for
the sample galaxies was obtained from the LEDA catalog\footnote{LEDA
database can be found at {\tt http://leda.univ-lyon1.fr}}.

\section{Collisionless Gas Models}

Can the gravitational potential of a BH mass induce gas motions which
explain $\sigma_{\rm gas}$? Current observations are consistent (e.g.,
Kormendy \& Gebhardt 2001, for a review) with most (and perhaps all)
early-type galaxies containing a black hole with a mass related to
$\sigma_{\rm star}$, i.e., the '$M_{\rm BH}- \sigma_{\rm star}$'
relation (Gebhardt et al. 2000; Ferrarese \& Merritt 2000). We
compute the expected $\sigma_{\rm gas}$ for such a BH mass under
idealized circumstances. We assume the gas is a collisionless
isotropic spherical distribution of gas cloudlets orbiting the BH. The
predicted gas velocity dispersion $\sigma_{\rm gas}$(sphere,1) takes
into account aperture size, PSF and instrumental line width.

\begin{figure}
\centering
\includegraphics[height=2.in]{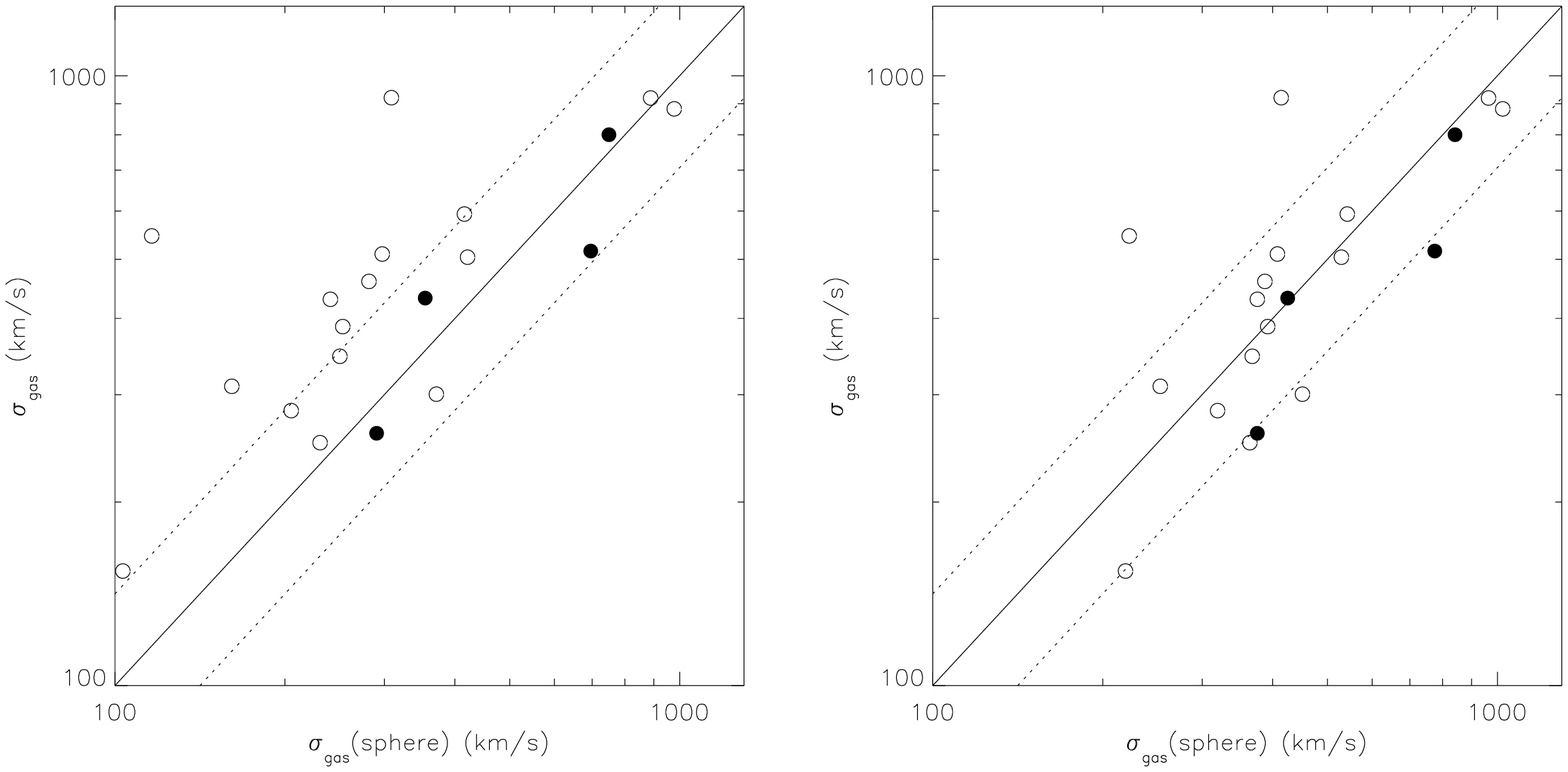}
\caption{The observed gas velocity dispersion $\sigma_{\rm gas}$ as a
function of predicted velocity dispersions for radio galaxies (open
circles) and relatively quiescent galaxies (filled circles). {\rm
(a):} The model for $\sigma_{\rm gas}$(sphere,1) assumes an isotropic
spherical distribution of gas cloudlets orbiting a BH with a mass according to the 
$M_{\rm BH} - \sigma_{\rm star}$ relation. {\rm (b):}
similar to (a) but now including the stellar mass distribution as well. 
The
dashed and solid lines indicate $\sigma_{\rm gas}$(sphere)=
$(1\sqrt{1/2},1,\sqrt{2}) \times \sigma_{\rm gas}$ respectively (to facilitate
comparison to the $M_{\rm BH} - \sigma_{\rm star}$). 
The figure suggests the gas
might be largely collisionless: see text for details.}
\label{figure2}
\end{figure}

Fig.~\ref{figure2}a shows that $\sigma_{\rm gas}$ typically exceeds $\sigma_{\rm
gas}$(sphere,1). However, the calculation neglects the stellar mass
present, which will increase $\sigma_{\rm gas}$(sphere,1).  To
illustrate the possible effect of this, Fig.~\ref{figure2}b shows $\sigma_{\rm
gas}$(sphere,2) assuming an isothermal spherical stellar mass
distribution in addition to the BH mass, i.e., $\sigma^2_{\rm gas}{\rm
(sphere,2)} = \sigma^2_{\rm gas}{\rm (sphere,1)} + \sigma^2_{\rm
star}$.  $\sigma^2_{\rm gas}$ typically agrees within a factor 2 to
$\sigma^2_{\rm gas}$(sphere,2). In other words, the $M_{\rm BH}$
needed to account for $\sigma^2_{\rm gas}$ agrees within a factor
$\sim 2$ with $M_{\rm BH}$ as predicted by the $M_{\rm BH} -
\sigma_{\rm star}$ relation (because $M_{\rm BH} \sim \sigma^2_{\rm
gas}$), under these simplifying assumptions. The estimated intrinsic
scatter in the relation is also a factor 2 (Tremaine et al. 2002). Thus, an
additional contribution to $\sigma_{\rm gas}$ from collisions (e.g.,
shocks/turbulence) is not required in this case.

\begin{figure}
\centering
\includegraphics[height=2.in]{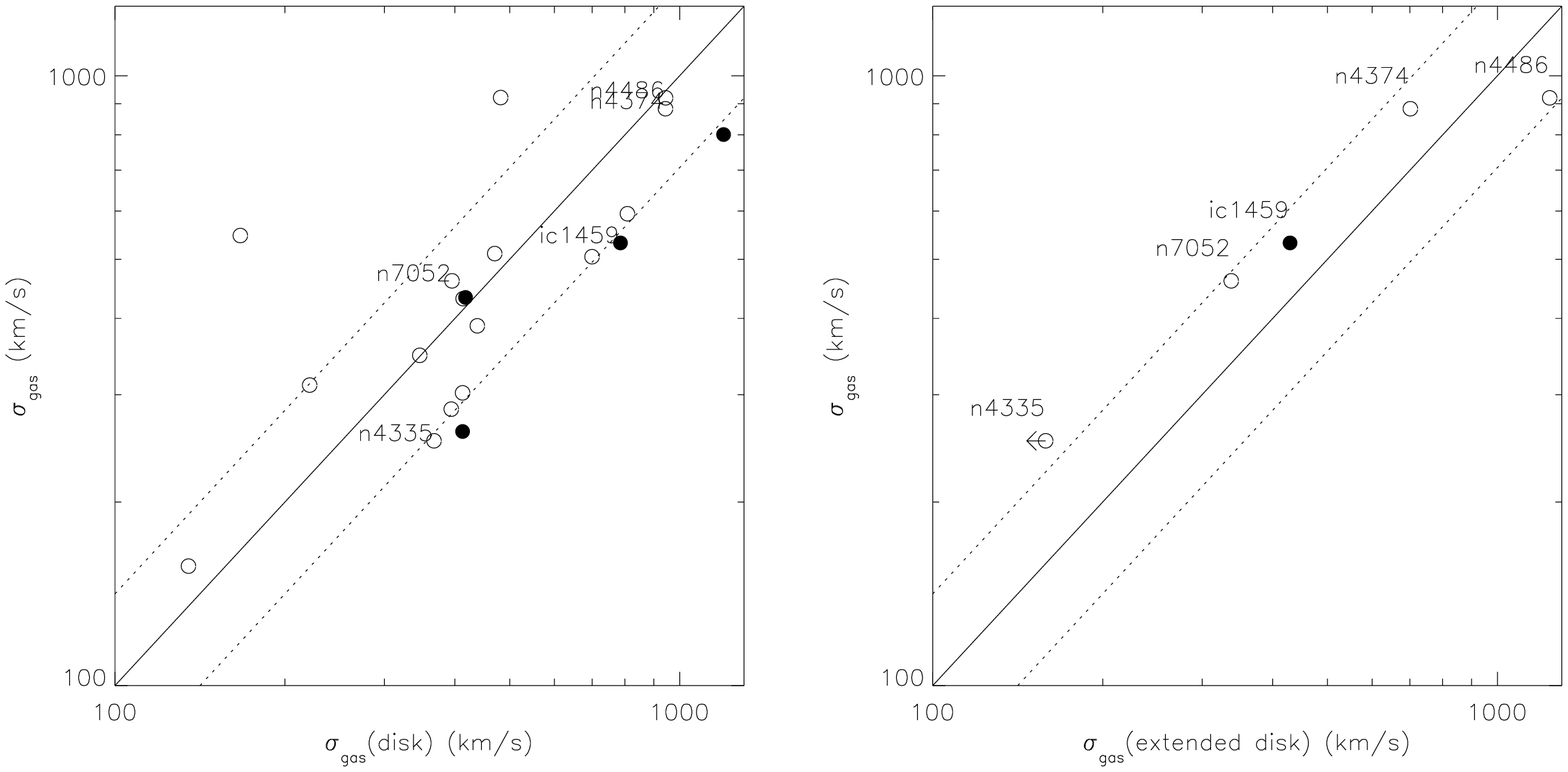}
\caption{The observed gas velocity dispersion $\sigma_{\rm gas}$ as a
function of predicted velocity dispersions for radio galaxies (open
circles) and relatively quiescent galaxies (filled circles). {\rm
(a):} The model assumes an infinitely thin gas disk in circular
rotation, inclined by $60^{\circ}$ to the line of sight and a BH mass
$M_{\rm BH}$ according to the $M_{\rm BH} - \sigma_{\rm star}$
relation. {\rm (b):} same as (a), but now for $M_{\rm BH}$ inferred
from published, detailed modeling of the gas mean velocities $v_{\rm
gas}$. The figure suggests that $v_{\rm gas}$ and $\sigma_{\rm gas}$
cannot be simultaneously fit by the gas disk model: see text for
further details. }
\label{figure3}
\end{figure}

Alternatively, we can assume that the collisional gas settles in a
disk. We idealize the disk as an infinitely flat disk of gas
particles/cloudlets, inclined by 60$^{\circ}$ to the line of sight,
and in circular rotation in the same combined BH and stellar potential
as used in the spherical case. On the nucleus, the differential
rotation will broaden the emission-line flux
profile. Fig.~\ref{figure3}a shows that the predicted gas velocity
dispersion $\sigma_{\rm gas}$(disk) roughly agrees with $\sigma_{\rm
gas}$. Hence no shocks/turbulence are required on average; a similar
result as for the gas spherical model.

Gas dynamical models of galactic nuclei, based on gas mean velocity
$v_{\rm gas}$ instead of $\sigma_{\rm gas}$ often reach a
different conclusion. For five sample galaxies, the central BH mass
has been measured directly from gas mean velocities $v_{\rm gas}$ in
extended gas disks. Fig.~\ref{figure3}b shows that the velocity dispersion
$\sigma_{\rm gas}(v_{\rm gas})$ predicted by our gas disk model for
{\it these} BH masses underpredicts $\sigma_{\rm gas}$ in four cases. Hence,
an additional contributor to the velocity dispersion is needed to
explain the observed values in our gas disk model. This agrees with
the mean velocity modeling which invokes an additional source of
dispersion, 'gas turbulence', to explain $\sigma_{\rm gas}$
(Maciejewski \& Binney 2001; van der Marel \& van den Bosch 1998;
Cappellari et al. 2002; Verdoes Kleijn et al. 2002). Only for NGC 4486
does a detailed gas disk modeling by Macchetto et al.~(1997) account
for both $v_{\rm gas}$ and $\sigma_{\rm gas}$, consistent with our
modeling. It is reassuring that our simplistic modeling is consistent with
the more detailed gas dynamical models on the difference between NGC
4486 and the other four galaxies.

It is less reassuring that Fig.~\ref{figure3}b suggests that a thin, circular gas
disk model in general cannot explain simultaneously observed gas
velocities and dispersions. For a simultaneous fit one has to invoke
gas turbulence. Alternatively, the extended gas might somehow rotate at
subcircular velocities, which would lead the models to infer a lower BH mass
and hence lower nuclear differential rotation.

Qualitatively, Fig.~\ref{figure2} suggests another possible solution to this
problem: the gas changes from a flat disk
configuration outside the nucleus to a (more) spherical distribution
closer to the nucleus. The nuclear spherical gas distribution explains
the central $\sigma_{\rm gas}$. Furthermore, the new configuration
implies that (i) the flat gas disk effectively has an inner radius and
(ii) only part of the observed nuclear emission-line flux is coming
from a disk. Both changes put the gas associated with the circular
disk at a larger mean distance from the BH, compared to the continous
disk model, and hence at lower circular velocities for the same BH mass.

\section{Discussion \& Conclusions}

The simple models discussed above suggest that the observed excess in
velocity dispersion in nuclear emission-line gas disks could be
explained by a more spherical gas distribution towards the nucleus. In
such a model, the gas remains collisionless (as in the customary
single gas disk model) and thus avoids invoking ad hoc turbulence or
shocks which require a permanent input of kinetic energy to
persist. Nevertheless, it could well be that activity associated with
the BH can provide such a continuous input of energy. The outliers in
Figs.~\ref{figure2} \&~\ref{figure3} and the tendency for the collisionless models to
underpredict $\sigma_{\rm gas}$ more for active galaxies rather than for
quiescent galaxies perhaps favor this scenario.

More sophisticated modeling and further observations can help to
determine between different gas configurations. On the one hand, more
accurate modeling of the stellar mass density and the emission-line
velocity profiles are expected to decrease the predicted $\sigma_{\rm
gas}$ at fixed BH mass. The central density profile of bright
ellipticals is generally shallower than an isothermal
distribution. Emission-lines often show extended wings: the true 2nd
moment is then larger than the Gaussian 2nd moment. This effect is
taken into account in the gas disk model, but not in the spherical
model.  On the other hand, emission-line flux profiles are typically
marginally resolved. They may contain a significant contribution from
an unresolved component which can be arbitrarily close to the BH. This
might increase the predicted $\sigma_{\rm gas}$.

UV spectroscopy can determine directly if shocks are present. In the
presence of shocks the gas can be shock-ionized. If absent, the
emission-line gas is only photo-ionized. The two ionization mechanisms
produce different UV emission-line ratios (e.g., Dopita et al. 1997).

In summary, the current results cannot distinguish between the various
gas configurations.  However, accurate BH mass measurements (and hence
BH demography) and BH accretion models depend on such a
distinction. Improvement of the simplistic gas dynamical modeling
presented here and UV observations can help to resolve this.

\end{document}